\documentclass{revtex4}

\usepackage{amsmath,amssymb}
\usepackage{graphicx}

\draft

\begin{document}
\titlepage
\title{Cosmological dynamics of D-cceleration}
\author{Xin-He Meng$^{1,2}$ \footnote{xhmeng@phys.nankai.edu.cn}
 \ \ Peng Wang$^1$ \footnote{pewang@eyou.com}
} \affiliation{1.  Department of Physics, Nankai University,
Tianjin 300071, P.R.China \\2. Department of Physics, University
of Arizona, Tucson, AZ 85721 USA}

\begin{abstract}
We study the cosmological dynamics of an effective theory for a
strongly coupled scalar field in the moduli space of
$\mathcal{N}=4$ supersymmetric Yang-Mills theory recently proposed
by Silverstein and Tong, called "D-cceleration". We discuss
various Energy Conditions in this theory. Then we prove the
inflationary attractor property using the Hamilton-Jacobi method
and study the phase portrait as well as the cosmological evolution
of the scalar field.
\end{abstract}

\maketitle

\section{Introduction}

It is a great challenge to implement inflation in string/M-theory.
Currently, to achieve this goal, there are mainly two types of
models: one is the modification of the Friedmann equation in the
braneworld scenario (see, e.g. \cite{brane} for a review); another
is specific model of inflaton field from string/M-theory (see,
\cite{Linde} for a recent important development). In this paper,
we will consider a scenario belonging to the second type recently
proposed by Silverstein and Tong \cite{Tong}.

In \cite{Tong}, Silverstein and Tong considered the inflaton as
the scalar field in the moduli space of the $\mathcal{N}=4$
supersymmetric Yang-Mills (SYM) theory. They call this model
"D-cceleration". In the strong coupling region, the effective
description of the theory is via gravity and string theory by
using the AdS/CFT correspondence \cite{AdSCFT}. Then in the
gravity side, the causal speed-limit in the bulk will translate
into a speed limit on moduli space. This has also been observed
before by Kabat and Lifschytz \cite{Kabat}. Interpreted in the
field theory side of the correspondence, this result reflects the
breakdown of the moduli-space $\sigma$-model approximation due to
the growing importance of the higher derivative terms. The
dynamics will be changed dramatically from the canonical scalar
field theory due to this speed limit and may lead to inflations
without a flat potential.

The effective action comes from, as written in field theory
variable, the Dirac-Born-Infeld action of a probe D3-brane moving
in AdS$_5\times$S$^5$ \cite{Tong, AdSCFT} with metric signature
$\{-,+,+,+\}$,
\begin{equation}
S=-\frac{1}{g_{YM}^2}\int d^4x\sqrt{-g}
[\frac{\phi^4}{\lambda}\sqrt{1+\frac{\lambda(\partial_\mu\phi)^2}{\phi^4}}+V(\phi)-\frac{\phi^4}{\lambda}],\label{1.1}
\end{equation}
where $g_{YM}$ is the Yang-Mills coupling, and $\lambda=Ng_{YM}^2$
is the 't Hooft coupling, in which $N$ is the rank of the SYM
theory. The string coupling $g_s$ is given by
$g_s=g_{YM}^2/(4\pi)$. It is worth commenting that in the pure
$\mathcal{N}=4$ SYM theory, there does not have the potential
$V(\phi)$ (The dynamics of the $\phi$ field without the potential
is analyzed in Sec.3.2 of \cite{Tong}). The potential $V(\phi)$
arises only after couple $\phi$ to other sectors involved in a
full string compactification. However, as we will show in
Sec.\ref{EC}, a simple argument based on Strong Energy Condition
shows that if we want the $\phi$ field to be a reasonable
candidate of inflaton, we must consider a non-zero potential.

It is interesting to note that when
$\lambda(\partial_\mu\phi)^2/\phi^4\ll 1$, expanding the action
(\ref{1.1}) to first order in
$\lambda(\partial_\mu\phi)^2/\phi^4$, it just reduces to the
lagrangian of a canonical scalar field theory
$\mathcal{L}={1\over2}(\partial_\mu\phi)^2+V(\phi)$. So the
lagrangian (\ref{1.1}) can be viewed as a correction to the
canonical scalar field theory when the kinetic term
$(\partial_\mu\phi)^2$ is large. On the other hand, the lagrangian
(\ref{1.1}) is also similar to the effective action of tachyon
field \cite{tachyon} $\mathcal{L}=V(T)\sqrt{1+(\partial_\mu
T)^2}$, which is a physical example of the general scenario of
"k-inflation" \cite{kinflation}. So it is also conceivable that
the dynamics of the theory (\ref{1.1}) will be similar in some
respects to the tachyon field. The author of \cite{Tong} mentioned
that the theory (\ref{1.1}) can be viewed as a physical example of
k-inflation. However, the general lagrangian of k-inflation is of
the form $\mathcal{L}=K(\phi)p(X)$, where
$X={1\over2}\partial_\mu\phi\partial^\mu\phi$, $K$ and $p$ are two
arbitrary functions. Then we can see that the lagrangian
(\ref{1.1}) actually does not belong to this form. So this
lagrangian is not an example of k-inflation in a precise meaning
and the analysis of the dynamics of k-inflation cannot be applied
to it.

The most notable feature of the lagrangian (\ref{1.1}) is that it
automatically imposes a restriction on $\dot{\phi}$, the speed
limit:
\begin{equation}
\dot\phi^2\leq\frac{\phi^4}{\lambda}.\label{1.2}
\end{equation}
Thus when $\phi$ is small, the $\phi$ field will be enforced to
roll slowly without the need of a flat potential.

The energy density $\rho$ and pressure $p$ of the $\phi$ field
follows from the action (\ref{1.1}) (Note that since we are
interested in the cosmological dynamics of the $\phi$ field, it is
reasonable to assume that $\phi$ is spatially homogeneous),
\begin{equation}
\rho=\frac{\phi^4}{\lambda\sqrt{1-\lambda\dot\phi^2/\phi^4}}+V(\phi)-\frac{\phi^4}{\lambda},\label{1.5}
\end{equation}
\begin{equation}
p=-{\phi^4\over
\lambda}\sqrt{1-\lambda\dot\phi^2/\phi^4}-V(\phi)+\frac{\phi^4}{\lambda},\label{1.6}
\end{equation}
and the equation of motion for $\phi$ can be obtained either by
varying $\phi$ in (\ref{1.1}) or by the energy conservation
equation $\dot{\rho}+3H(\rho+p)=0$,
\begin{equation}
\ddot\phi-\frac{6\dot\phi^2}{\phi}+\frac{4\phi^3}{\lambda}+3H(1-\frac{\lambda\dot\phi^2}{\phi^4})\dot\phi
+(V'(\phi)-\frac{4\phi^3}{\lambda})(1-\frac{\lambda\dot\phi^2}{\phi^4})^{2/3}=0\label{2.6}
\end{equation}
It is easy to check that
\begin{equation}
\phi=\frac{\sqrt\lambda}{t+\sqrt\lambda/\phi_i}\label{1.31}
\end{equation}
where $\phi_i$ is the initial value, is always a solution of the
evolution equation (\ref{2.6}) which satisfies the speed limit
exactly, i.e. $\dot\phi^2=\frac{\phi^4}{\lambda}$, irrespective of
the precise form of the potential $V$. However, this cannot be a
stable solution: from the expression for the energy density
$\rho$, i.e. Eq.(\ref{1.5}), we can see that when $\phi$ satisfies
the speed limit exactly, the energy density tends to infinity
(This also has an interesting dual interpretation from AdS/CFT:
when the speed of the D-brane tends to unity, according to
relativity, its energy tends to infinity). This means that the
action (\ref{1.1}) is not sensible when $\phi$ satisfies the speed
limit exactly.

\section{Energy conditions}
\label{EC}

When facing a new field theory that intended to apply to
cosmology, one natural question one would like to ask is: under
what conditions it will satisfy or violate the various Energy
Conditions \cite{hawking}?

First, it is easy to see from Eqs.(\ref{1.5}), (\ref{1.6}) that
$\phi$ always satisfies the Weak Energy Condition: $\rho\ge0$ and
$\rho+p\ge0$. This implies that the equation of state of $\phi$
$\omega\equiv p/\rho$ always satisfies $\omega\ge-1$. Thus this
theory will not suffer from the possible instability of vacuum
\cite{carroll-phantom} that might cause problems in a phantom
field theory which is characterizes by $\omega<-1$.

Second, with a little more labor, it is also easy to say that the
Dominant Energy Condition: $\rho\ge|p|$, is also always satisfied
by $\phi$. Thus, from the vacuum conservation theorem of Hawking
and Ellis \cite{hawking}, (see also \cite{carter} a recent review
and a simplified proof), the energy of $\phi$ field cannot
propagate outside the light-cone.

At last, let's come to the Strong Energy Condition: $\rho+p\ge0$
and $\rho+3p\ge0$. From Eqs.(\ref{1.5}), (\ref{1.6}),
\begin{equation}
\rho+3p={2\phi^4\over\lambda}(1-\sqrt{1-\lambda\dot{\phi}^2/\phi^4})+{\dot{\phi}^2\over
\sqrt{1-\lambda\dot{\phi}^2/\phi^4}}-2V(\phi). \label{2.00}
\end{equation}
Then it is easy to see that when $V=0$, $\phi$ will always satisfy
the Strong Energy Condition. As is well-known, for a matter to
drive an accelerated universe, it is just necessary and sufficient
for that matter to violate the Strong Energy Condition. So if
$V=0$, as in the case of pure $\mathcal{N}=4$ SYM theory, $\phi$
field cannot be a viable candidate of infaton. Whether $\phi$
field can violate the Strong Energy Condition depends on the
precise form of the potential. In this sense, the $\phi$ field is
more like a canonical scalar field rather than a tachyon field.
Since in the latter case, whether the tachyon field can violate
the Strong Energy Condition depends only on the value of
$|\dot{T}|$ (see, e.g. \cite{gibbons}).

\section{Inflationary attractor property}
\label{IAP}

Inflation can be predictive only if the solution exhibits an
attractor behavior, where the differences between solutions of
different initial conditions rapidly vanish (see Sec. 3.7 of
\cite{Liddle} for a review). Thus this section is devoted to the
study of the inflationary attractor property of the $\phi$ field
for a general potential $V(\phi)$.

The Friedmann equation reads,
\begin{equation}
H^2=\frac{1}{3g_sM_p^2}\rho\label{2.5}
\end{equation}
where $H=\dot a/a$ is the Hubble parameter.

We will use the Hamilton-Jacobi formulation for the Modified
Friedmann equation \cite{HJ}, which is a powerful and very
effective tool for analyzing the inflationary attractor property
\cite{attractor}. This formulation is also very useful in
obtaining solutions of the evolution equations \cite{Tong}. In
this formulation, we will view the scalar field $\phi$ as the time
variable and this requires that the $\phi$ field does not change
sign during inflation. Without loss of generality, we can choose
$\dot\phi>0$ in the following discussions. If this is not
satisfied, it can be brought about by redefining
$\phi\rightarrow-\phi$.

The Hamilton-Jacobi formulation of the Friedmann equation reads
\cite{Tong}
\begin{equation}
V(\phi)=3(g_sM_p^2)H(\phi)^2-\frac{(g_sM_p^2)\phi^4}{\lambda}\sqrt{1/(g_s^2M_p^4)+4\lambda
H'(\phi)^2/\phi^4}+\frac{\phi^4}{\lambda}\label{2.7}
\end{equation}
and
\begin{equation}
\dot\phi=\frac{-2H'(\phi)}{\sqrt{1/(g_s^2M_p^4)+4\lambda
H'(\phi)^2/\phi^4}}\label{2.71}
\end{equation}

Supposing $H_0(\phi)$ is any solution to Eq.(\ref{2.7}), which can
be either inflationary or non-inflationary. We consider a small
homogeneous perturbation $\delta H(\phi)$ to this solution. The
attractor property will be satisfied if it becomes smaller as
$\phi$ increases. Substituting $H(\phi)=H_0(\phi)+\delta H(\phi)$
into Eq.(\ref{2.7}) and linearizing, we find that the perturbation
obeys
\begin{equation}
2H_0'\delta H'=3H_0\sqrt{1/(g_s^2M_p^4)+4\lambda
H'(\phi)^2/\phi^4}\delta H\label{2.8}
\end{equation}
which has had the general solution
\begin{equation}
\delta H(\phi)=\delta
H(\phi_i)\exp[\int^\phi_{\phi_i}3H_0\sqrt{1/(g_s^2M_p^4)+4\lambda
H'(\phi)^2/\phi^4}\frac{d\phi}{2H_0'}]\label{2.9}
\end{equation}
where $\delta H(\phi_i)$ is the value at some initial point
$\phi_i$. From Eq.(\ref{2.71}), since $H_0'$ and $d \phi$ have the
opposite sign, if $H_0$ is an inflationary solution, all linear
perturbations are damped at least exponentially.

Thus we conclude that the inflationary attractor property holds
for the field $\phi$. See Fig.1 for the phase portrait of the
potential (\ref{1.4}) where we can clearly perceive this
conclusion.

\section{Phase portrait and cosmological evolution}
\label{PP}

In this section, we will study numerically the dynamics of the
$\phi$ field for the simplest potential that may lead to a power
law acceleration \cite{Tong}
\begin{equation}
V(\phi)=\frac{1}{2}m^2\phi^2\label{1.4}
\end{equation}
It has been shown in \cite{Tong} that at late time, $\phi$ will
just tend to solution (\ref{1.31}). We choose this potential is
because that in addition to its simplicity, as argued in
\cite{Tong}, higher power terms in the effective potential will be
small compared to the quadratic term without significant tuning.

However, the lagrangian (\ref{1.1}) cannot be trusted for
arbitrarily small $\phi$ because of the back reaction of the probe
brane on the geometry. It is shown in \cite{Tong} that the theory
can be trusted only for $\phi$ greater than a critical value
\cite{Tong},
\begin{equation}
\phi^2>\frac{1}{2}g_sm^2\equiv\phi_c^2\label{1.7}
\end{equation}

Also in \cite{Tong}, it has been argued that the potential
(\ref{1.4}) can drive a power law inflation only if $m$ satisfies
\begin{equation}
\frac{m^2}{M_p^2}>\frac{2g_s}{\lambda}\label{1.8}
\end{equation}
where the Planck scale $M_p^2=(8\pi G)^{-1}$. We have verified
numerically that this is indeed the case. Thus in the following
discussions, we will impose this bound on the parameter $m$.

In the following discussion, we will fix the tension parameter
$\lambda=100$ throughout for illustrative purposes.

To study the evolution, it is convenient to rewrite the evolution
equation (\ref{2.6}) as a set of two first order equations with
two independent re-scaled variables $Y_1=\phi/M$ and
$Y_2=\dot\phi/M^2$ where $M$ is the inflationary energy scale.
Then in terms of $Y_1$ and $Y_2$, the speed limit (\ref{1.2}) and
the bound (\ref{1.7}) translate into
\begin{equation}
Y_2^2\leq\frac{Y_1^4}{\lambda}\label{3.0}
\end{equation}
and
\begin{equation}
Y_1^2>\frac{1}{2}g_sr_m^2\equiv Y_{1c}\label{3.1}
\end{equation}
where $r_m\equiv m/M<1$ for any reasonable value of the scalar
field's mass $m$.

We define a dimensionless parameter$r_i=M/M_p$ for describing the
energy scale fraction. In the following discussion, we will always
take $r_i=10^{-3}$, which corresponds to the GUT scale for the M.
Then the evolution equation (\ref{2.6}) can be written as an
dynamically autonomous system:
\begin{equation}
Y_1'=\frac{Y_2}{H/M}\label{3.2}
\end{equation}
\begin{equation}
Y_2'=\frac{1}{H/M}[\frac{6Y_2^2}{Y_1}-\frac{4Y_1^3}{\lambda}-3\frac{H}{M}(1-\frac{\lambda
Y_2^2}{Y_1^4})
Y_2-(r_m^2Y_1-\frac{4Y_1^3}{\lambda})(1-\frac{\lambda
Y_2^2}{Y_1^4})^{3/2}]\label{3.3}
\end{equation}
where a prime denotes differentiation with respect to the number
of e-folding $N$ defined by
\begin{equation}
N=\ln\frac{a}{a_i}=\int^t_{t_i}Hdt,\label{3.4}
\end{equation}
thus $dN=Hdt=d\ln a$.

And in terms of $Y_1$ and $Y_2$, the Friedmann equation
(\ref{2.5}) can be rewritten as
\begin{equation}
(\frac{H}{M})^2=\frac{r_i^2}{3g_s}[\frac{Y_1^4}{\lambda\sqrt{1-\lambda
Y_2^2/Y_1^4}}+\frac{1}{2}r_m^2Y_1^2-
\frac{Y_1^4}{\lambda}]\label{3.5}
\end{equation}

Also in terms of $r_m$ and $r_i$, the bound (\ref{1.8}) translates
into
\begin{equation}
r_m>\sqrt{\frac{2g_s}{r_i^2\lambda}}\label{3.6}
\end{equation}

\begin{figure}
  \includegraphics[width=0.4\columnwidth]{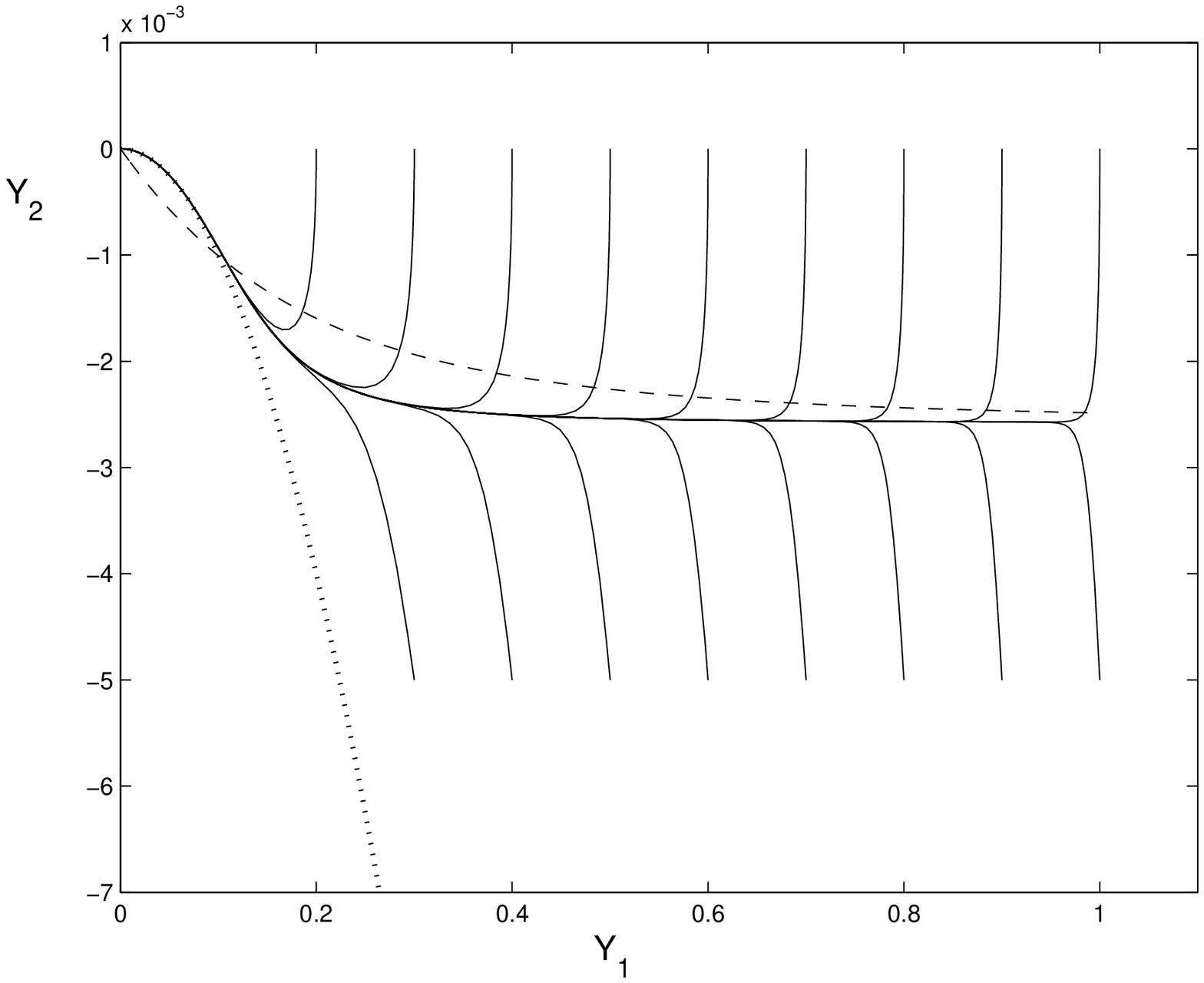}
  \includegraphics[width=0.4\columnwidth]{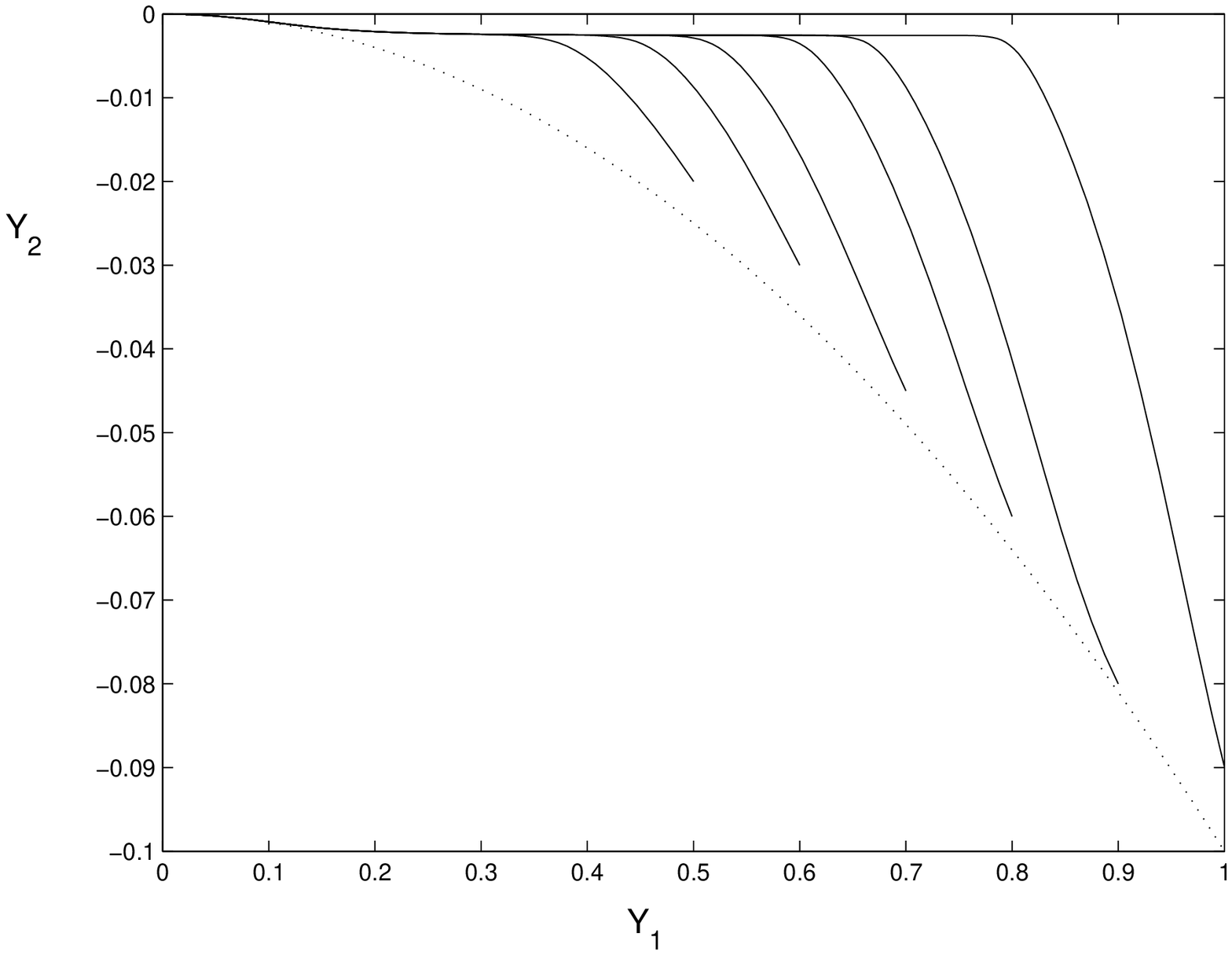}
  \caption{Phase portrait for the field $\phi$ in the re-scaled coordinates $(Y_1,Y_2)$ for $g_s=1\times 10^{-8}$ and
  $r_m=(10g_s/(r_i^2\lambda))^{1/2}$. The left panel show the behavior for small initial value of $Y_2$. The right panel
  shows the behavior for initial value of $Y_2$ close to the speed limit bound. In both panels,
  the dotted line corresponds to the curve that satisfies the speed limit exactly, i.e.
  $Y_2=-Y_1^2/\sqrt \lambda$. The dashed line in the right panel corresponds to the slow-roll approximated equation
  (\ref{3.9}).}\label{1}
\end{figure}

In the following we start the numerical analysis of the dynamical
system. Fig.1 shows the trajectories in the phase plane $(Y_1,
Y_2)$ with $g_s=1\times10^{-8}$ and
$r_m^2=10g_s/(r_i^2\lambda)=10^{-3}$. $Y_1$ will take values in
the region $Y_{1c}<Y_1<1$; $Y_2$ will take values in the region
$-Y_1^2/\sqrt{\lambda}<Y_2<Y_1^2/\sqrt{\lambda}$. We can see that
there is a curve that attracts all the trajectories which
correspond to the slow-roll curve. This confirms our conclusion in
the last section for the specific potential (\ref{1.4}). For all
the allowed initial value, $\phi$ tends to the slow-roll curve
quite soon after it begins to evolve. For simplicity, let us then
analyze in detail the dynamics of $\phi$ for small initial $Y_2$
that satisfies $\lambda Y_{2i}^2/Y_{1i}^4\ll 1$. Since
$Y_{1i}\sim1$ and $\lambda=100$, this corresponds to
$Y_{2i}\ll10^{-1}$.

See the left panel of Fig.1 where we have taken
$Y_{2i}\sim10^{-3}$. By the numerical calculations, the dynamics
of $\phi$ can be divided qualitatively into three stages. First,
when it begins to evolve, it quickly evolves to the slow-roll
curve. Second, it undergoes a slow-roll stage until $Y_1\sim0.2$.
Third, it evolves as the speed limit curve.

In the first stage, from the left panel of Fig.1, we can
consistently set $Y_1=const.$ and because we have assumed a small
initial $Y_2$, we can set $1-\lambda Y_2^2/Y_1^4\approx 1$. Thus
from Eq.(\ref{3.5}), $H/M=const.$ and $(H/M)^2\sim10^{-1}Y_1^2$.
Because $Y_2\sim10^{-3}$ and $Y_1\sim1$ in this stage, we have
$Y_2^2/(HY_1/M)\ll-Y_2$ with $Y_2<0$. Thus Eq.(\ref{3.3}) reduces
to
\begin{equation}
Y_2'=-3Y_2-\frac{r_m^2}{H/M}Y_1\label{3.61}
\end{equation}
which can be integrated to give
\begin{equation}
Y_2=(Y_{2i}+\frac{r_m^2}{3H/M}Y_1)\exp(-3N)-\frac{r_m^2}{3H/M}Y_1\label{3.62}
\end{equation}
Thus in this region, $Y_2$ is damped exponentially to the
slow-roll curve with $Y_1$ kept almost constant as we can see that
from the upper right of the left panel. Furthermore, we can see
that during the slow-roll stage, $Y_2\sim-r_m^2/(3H/M)$.

In the second stage, since $\lambda Y_2^2/Y_1^4\sim 10^{-3}$, we
may set $1-\lambda Y_2^2/Y_1^4\approx1$ and $Y_2'\approx0$ in the
dynamical system as described by Eqs. (\ref{3.2}), (\ref{3.3}),
and (\ref{3.5}). Then the system can be approximated by
\begin{equation}
\frac{6Y_2^2}{Y_1}-3\frac{H}{M} Y_2-r_m^2Y_1=0\label{3.7}
\end{equation}
and
\begin{equation}
(\frac{H}{M})^2=\frac{r_i^2}{6g_s}r_m^2Y_1^2\label{3.8}
\end{equation}

This is a simple algebraic system of $(Y_1, Y_2)$ and can be
solved exactly to give
\begin{equation}
Y_2=[3\frac{H}{M}-\sqrt{9(\frac{H}{M})^2+24r_m^2}]\frac{Y_1}{12}\label{3.9}
\end{equation}
The dashed line in the left panel of Fig.1 corresponds to the
phase portrait of this slow-roll approximated evolution equation.
Furthermore, since $(H/M)^2\sim10^{-1}Y_1^2$ and $r_m^2=10^{-3}$,
Eq.(\ref{3.9}) can further reduce to
\begin{equation}
Y_2=-\frac{r_m^2}{3H/M}Y_1\label{3.10}
\end{equation}
This is consistent with Eq.(\ref{3.62}).

In the third stage, when $Y_{1c}<Y_1<0.2$, the attractor curve
coincides with the limit curve that satisfies the speed limit
exactly, i.e. $Y_2=-Y_1^2/\sqrt \lambda$. This has already been
pointed out in \cite{Tong} as the late time behavior of $\phi$ by
using the Hamilton-Jacobi method. However, as we have showed,
$\phi$ will behave in this way well before it reaches the critical
point $\phi_c$.

\section{Conclusions and Discussions}
\label{CD}

In this paper, we studied the cosmological dynamics of the
D-cceleration scenario. We proved the inflationary attractor
property using the Hamilton-Jacobi method and studied the phase
portrait. Thus in this respect, this model can be a reasonable
model of inflation. Of course, there are still many works needed
to be done such as confrontation with observation to see whether
this is a viable model of inflation.

It is interesting to ask whether the theory (\ref{1.1}) can also
be a viable candidate for quintessence which can drive an
acceleration in the current universe (see, e.g. \cite{carroll-de}
for a recent review). This is interesting because the model
building of quintessence faces the same problem as inflation: it
requires the quintessence field to be in a slow-roll phase which
requires a flat potential. But this is rather difficult to achieve
from a particle physics point of view. So the important feature of
D-cceleration, i.e. slow-roll without a flat potential, is also
very attractive as a quintessence candidate. However, the problem
is that, assuming we still work with the potential (\ref{1.4}),
the theory will make sense only under the condition that
$\phi>\phi_c$. So if we want to apply this model to the late
universe, we must check that $\phi$ will satisfy $\phi>\phi_c$
after inflation for a wide range of initial conditions. Although
we still cannot present a definite answer to this question,
through numerical computation, we found that the evolution of
$\phi$ will be extremely slowed down before it reaches $\phi_c$.
So this possibility still remains.

Following the same line of reasoning after the effective theory of
tachyon is proposed, it is also interesting to consider the
evolution of the D-cceleration in brane-world scenarios such as
the well-studied Randall-Sundrum II model \cite{tachyonbrane}.

\section*{Acknowledgements}

We would like to thank D.Tong for reading the manuscript and
helpful comments. We also would like to thank D.Lyth, S.Odintsov,
X.P.Wu and X.M.Zhang for helpful correspondence. This work is
supported partly by ICSC-World Laboratory Scholarship, China NSF
and Doctoral Foundation of National Education Ministry.


\begin{thebibliography}{99}

\bibitem{brane}
J. E. Lidsey, astro-ph/0305528; J. E. Lidsey, D. Wands and E. J.
Copeland, Phys. Rep. \textbf{337} (2000) 343; M. Gasperini and G.
Veneziano, Phys. Rep. \textbf{373} (2003) 1; M. Quevedo, Class.
Quant. Grav. \textbf{19} (2002) 5721 [hep-th/0210292].

\bibitem{Linde}
S. Kachru, R. Kallosh, A. Linde, J. Maldacena, L. McAllister and
S. P. Trivedi, JCAP \textbf{0310} (2003) 013 [hep-th/0308055].

\bibitem{Tong}
E. Silverstein and D. Tong, hep-th/0310221.

\bibitem{Kabat}
D. Kabat and G. Lifschytz, JHEP \textbf{9905} (1999) 005
[hep-th/9902073].

\bibitem{AdSCFT}
O.~Aharony, S.~S.~Gubser, J.~M.~Maldacena, H.~Ooguri and Y.~Oz,
Phys. Rept. \textbf{323} (2000) 183 [hep-th/9905111].

\bibitem{tachyon}
A. Sen, JHEP \textbf{0204} (2002) 048; ibid, JHEP \textbf{0207}
(2002) 065; M. R. Garousi, Nucl. Phys. \textbf{B584} (2000)
284-299 [hep-th/0003122].


\bibitem{kinflation}
C.~Armendariz-Picon, T.~Damour and V.~Mukhanov, Phys. Lett.
\textbf{B456} (1999) 209 [hep-th/9904075];

\bibitem{hawking} S.~W.~Hawking and G.~F.~R.~Ellis, {\it The Large
Scale Structure of Space Time} (Cambridge U.P., 1973).

\bibitem{carroll-phantom}
S.~M.~Carroll, M.~Hoffman and M.~Trodden, Phys. Rev. {\bf D68}
(2003) 023509 [astro-ph/0301273].

\bibitem{carter}
B. Carter, gr-qc/0205010.

\bibitem{gibbons}
G.~W.~Gibbons, Phys. Lett. {\bf B537} (2002) 1 [hep-th/0204008].

\bibitem{Liddle} A. R. Lidde and D. H. Lyth, {\it Cosmological Inflation and Large
Scale Structure}, Cambrigde University Press, 2000;

\bibitem{HJ} D.~S.~Salopek and J.~R.~Bond, Phys. Rev. \textbf{D42}
(1990) 3936; A. G. Muslimov, Class. Quant. Grav. \textbf{7} (1990)
231; J. E. Lidsey, Phys. Lett. \textbf{B273} (1991) 42; A. R.
Liddle, P. Parsons and J. D. Barrow, Phys. Rev. \textbf{D50}
(1994) 7222 [astro-ph/9408015];

\bibitem{attractor}
Z. K. Guo, H. S. Zhang and Y. Z. Zhang, hep-ph/0309163; Zong-Kuan
Guo, Y. S. Piao, R. G. Cai and Y. Z. Zhang, Phys. Rev.
\textbf{D68} (2003) 043508 [hep-ph/0304236]; X. H. Meng and P.
Wang, Class. Quant. Grav. {\bf 21} (2004) [hep-ph/0312113];

\bibitem{carroll-de}
S.~M.~Carroll, astro-ph/0310342.

\bibitem{tachyonbrane}
M.~C.~Bento, O.~Bertolami, A.~A.~Sen, Phys.\ Rev.\ {\bf D67}
(2003) 063511 [hep-th/0208124].

\end{thebibliography}
\end{document}